\documentclass[pdflatex,sn-mathphys-num]{sn-jnl}% Math and Physical Sciences Numbered Reference Style 
\usepackage{graphicx}%
\usepackage{multirow}%
\usepackage{amsmath,amssymb,amsfonts}%
\usepackage{amsthm}%
\usepackage{mathrsfs}%
\usepackage[title]{appendix}%
\usepackage{xcolor}%
\usepackage{textcomp}%
\usepackage{manyfoot}%
\usepackage{booktabs}%
\usepackage{algorithm}%
\usepackage{algorithmicx}%
\usepackage{algpseudocode}%
\usepackage{listings}%

\usepackage{moreverb,url}

\newcommand\BibTeX{{\rmfamily B\kern-.05em \textsc{i\kern-.025em b}\kern-.08em
		T\kern-.1667em\lower.7ex\hbox{E}\kern-.125emX}}

\usepackage{amsmath,amssymb,mathtools,centernot}
\newcommand\independent{\protect\mathpalette{\protect\independenT}{\perp}}
\def\independenT#1#2{\mathrel{\rlap{$#1#2$}\mkern3mu{#1#2}}}

\usepackage{float}
\usepackage{bm}
\usepackage{times}
\usepackage{calc}
\usepackage{graphicx}

\newtheorem{assumption}{Assumption}
\usepackage{textgreek}
\usepackage{booktabs}
\usepackage{centernot}

\usepackage{enumitem}
\newlist{Properties}{enumerate}{2}
\setlist[Properties]{label=Property \arabic*., font=\textbf, itemindent=*}

\theoremstyle{thmstyleone}%
\theoremstyle{thmstyletwo}%

\theoremstyle{thmstylethree}%

\begin{document}

\title[Propensity Score Matching: To Use or Not to Use?]{Propensity Score Matching: Should We Use It in Designing Observational Studies?}


\author*[1]{\fnm{Fei} \sur{Wan}}\email{wan.fei@wustl.edu}

\affil*[1]{\orgdiv{Division of Public Health Sciences}, \orgname{Washington University in St Louis}, \orgaddress{\street{660 S. Euclid Ave}, \city{St Louis}, \postcode{63110}, \state{MO}, \country{USA}}}


\abstract{\textbf{Background:} Propensity Score Matching (PSM) stands as a widely embraced method in comparative effectiveness research. PSM crafts matched datasets, mimicking some attributes of randomized designs, from observational data. In a valid PSM design where all baseline confounders are measured and matched, the confounders would be balanced, allowing the treatment status to be considered as if it were randomly assigned. Nevertheless, recent research has unveiled a different facet of PSM, termed ``the PSM paradox." As PSM approaches exact matching by progressively pruning matched sets in order of decreasing propensity score distance, it can paradoxically lead to greater covariate imbalance, heightened model dependence, and increased bias, contrary to its intended purpose. \textbf{Methods:} We used analytic formula, simulation, and literature to demonstrate that this paradox stems from the misuse of metrics for assessing chance imbalance and bias. \textbf{Results:} Firstly, matched pairs typically exhibit different covariate values despite having identical propensity scores. However, this disparity represents a ``chance" difference and will average to zero over a large number of matched pairs. Common distance metrics cannot capture this ``chance" nature in covariate imbalance, instead reflecting increasing variability in chance imbalance as units are pruned and the sample size diminishes. Secondly, the largest estimate among numerous fitted models, because of uncertainty among researchers over the correct model, was used to determine statistical bias. This cherry-picking procedure ignores the most significant benefit of matching design-reducing model dependence based on its robustness against model misspecification bias. \textbf{Conclusions:} We conclude that the PSM paradox is not a legitimate concern and should not stop researchers from using PSM designs.}




\maketitle

\section{Background}
\label{sec:intro}

Propensity Score Matching (PSM) stands out as one of the most well-established and widely used strategies for exploring the comparative effectiveness of competing interventions in observational studies, such as comparing active treatment (referred to as ``treated") to placebo control (referred to as ``untreated") \cite{pearl_3._2010}. The propensity score (PS) represents the probability that a subject will receive the active treatment, given their baseline covariates \cite{rosenbaum_central_1983}. Throughout this discussion, we assume that all baseline variables are measured, which is crucial for a valid PSM design. By capitalizing on the ignorability and balancing properties of PS, matching subjects based on their exact propensity scores ensures that the distribution of baseline variables becomes identical between treated and untreated individuals, and the treatment assignment for matched subjects can be regarded as essentially random. PSM effectively enables us to extract an approximate randomized experiment from observational data, facilitating robust comparative analyses.

While numerous studies have meticulously examined and validated the properties of PSM through simulations and theoretical investigations \cite{austin_methods_2009, austin_comparison_2014, dehejia_causal_1999, dehejia_propensity_2002, zhao_using_2004, wan_matched_2019, abadie_matching_2016}, a more recent study by King and Nielson presents a contrasting view on PSM \cite{king_why_2019}. This study contends that PSM should not be used, as it paradoxically ``increases imbalance, inefficiency, model dependence, research discretion, and statistical bias at some point in both real-world data and data tailored to adhere to PSM theory," a phenomenon they term the ``PSM paradox." In the context of a one-to-one PSM design, the study illustrates that as study subjects were progressively pruned (e.g., by reducing the caliper size) and PSM was approaching exact matching, there was an initial improvement in balance. This progress continued until a specific point was reached, where covariates became nearly balanced, and PSM approximated a completely randomized design (CRD). However, it was at this juncture that the PSM paradox manifested itself, leading to a subsequent deterioration in balance, ultimately resulting in an increased bias in the effect estimation. Their findings have prompted researchers from various fields to cast widespread doubt on the validity of PSM \cite{wyler_von_ballmoos_commentary:_2021,sceats_nonoperative_2019,sareen_preoperative_2023}.

Nevertheless, there are certain issues within their study that challenge the validity of their recommendation:
\\
\vspace{0.1cm}
{\it{First}}, it remains unclear why there is a need to further narrow the caliper width, potentially leading to the exclusion of valuable data, especially when baseline characteristics are already balanced using a reasonably sized caliper. For example, previous research has shown that using a caliper width of 0.2 times the standard deviation of logit PS can effectively eliminate over 90\% of confounding bias \cite{austin_optimal_2011}. 
\\
\vspace{0.1cm}
{\it{Second}}, the study chose the sample average treatment effect (SATE) as the causal interest. While SATE represents the treatment effect for the specific study sample, it may diverge significantly from the population average treatment effect (PATE), especially when individual treatment effects exhibit heterogeneity, and the study sample is not randomly selected from the target population. Our primary interest often lies in estimating population-level quantities \cite{austin_methods_2009, austin_comparison_2014}. 
\\
\vspace{0.1cm}
{\it{Third}}, the chosen imbalance metric, which calculates the average ``pairwise" absolute distance in covariate space from each treated subject to the closest untreated unit, raises concerns. This metric's limitations are twofold: i) Even when propensity scores are identical, treated and untreated subjects in a matched pair typically exhibit covariate mismatches-a point already extensively discussed by Rosenbaum \cite{rosenbaum_design_2020}. ii) These mismatches occur randomly, with negative and positive mismatches having a similar occurrence. As the number of matched pairs increases, the distributions of matched covariates eventually become similar between the treated and untreated groups, and the between-group imbalance becomes negligible, despite individual units within each matched pair having different covariate values. These are key implications of the balancing score and the strongly ignorability properties of PS. Common multivariate distance metrics, such as the ``mahalanobis" distance, cannot capture the ``chance" nature of covariate imbalance. 
\\
\vspace{0.1cm}
{\it{Lastly}}, Since the balancing score and the strongly ignorability properties of PS ensure that any mismatches in covariate values between matched pairs are random occurrences and don't lead to residual confounding, these mismatches can not bias the effect estimation. To demonstrate the increased statistical biases in the PSM paradox, the study adopts a biased approach for effect estimation, which involves the use of a well-known biased estimator: selecting the largest estimate from among hundreds of competing linear models. This rationale is based on the assumption that researchers may be uncertain about the correct model and should explore all possible models in post-matching analysis, potentially cherry-picking the best estimate. However, this approach neglects one of the most significant benefits of the matching design - its resilience against model misspecification. Importantly, this bias doesn't stem from imbalances in confounders between comparison groups.

In light of all these uncertainties, our study aims to determine whether the PSM paradox warrants legitimate concern, and whether researchers should avoid using PSM in comparative effectiveness research or not.   

\section{Methods}
\label{sec2}

\subsection{Definitions and assumptions}
The Rubin Causal Model (RCM), introduced by Rubin in 1974, is a widely used framework for defining causal effects \cite{rubin_estimating_1974}. In this model, we denote the binary treatment status as $A$, where 1 represents the treatment of interest, and 0 represents a control condition. Additionally, let $\bm{X}$ represent a vector of $p$ confounders at baseline, and $Y$ denote a continuous outcome variable. $Y(a)$ denote the potential outcome under the treatment $a$, where $a\in \{0,1\}$. We further make the following assumptions:
\begin{assumption}
	The Stable Unit Treatment Value Assumption (SUTVA), which consists of two sub-assumptions: 
	\begin{itemize}
		\item[(i)] The potential outcomes for any unit do not vary with the treatment assigned to other units (no interference between units).
		\item[(ii)] For each unit, there are no different versions of each treatment level (no hidden versions of treatments).
	\end{itemize} 
\end{assumption}
Under SUTVA, the potential outcomes of each individual $i$ depends only on the treatment assigned to this unit, not the treatments assigned to other units. For each individual $i$, $Y_i(1)$  represents the potential outcome that would have been observed if individual $i$ received the treatment of interest, while $Y_i(0)$  represents the potential outcome if individual $i$ received the control. The observed outcome is denoted as $Y_i=A_i Y_i(1)+(1-A_i)Y_i(0)$.
\begin{assumption}
	The conditional ignorable treatment assignment assumption
	\begin{align*}
		\{Y_i(1),Y_i(0)\} \independent A_i | \bm{X}_i
	\end{align*}
	Conditional on observed confounders, the treatment status can be considered as randomly assigned.
\end{assumption}
We also impose the Positivity assumption as follows:
\begin{assumption}
	For all observed covariates $\bm{x}$ where $P(\bm{x})>0$, $0<P(A=1|\bm{X}=\bm{x})<1$
\end{assumption}
This is also known as the common support or overlap assumption because it entails that the conditional distributions $P(A = 1 |\bm{X}=\bm{x}) $ and $P(A = 0 |\bm{X}=\bm{x})$ must share a common support.

\subsection{Population and sample causal estimand}
It's important to note that for any given individual $i$, only one potential outcome from the pair  $\{Y_i(0), Y_i(1)\}$  can be observed. As a result, individual-level treatment effects $Y_i(1)-Y_i(0)$ cannot be identified, leading researchers to often focus on average treatment effects (ATE). In the literature, there are two types of ATEs: one at the population level and the other at the sample level.

We assume there is a population of $N$ units in the super-population, from which we draw a sample $S$ with $n$ individuals. The population average treatment effect (PATE) is defined as
\begin{align*}
	\tau_{PATE}&=\mathbb{E}(Y(1)-Y(0)) \\
	&=\frac{1}{N}\sum^N_{i=1} \left( Y_i(1)-Y_i(0) \right)
\end{align*}
which is the difference in potential outcomes averaged across the $N$ units in the super-population. The sample average treatment effect (SATE) for our sample $S$ is defined as 
\begin{align*}
	\tau_{SATE}&=\mathbb{E}(Y(1)-Y(0)|S=1) \\
	&=\frac{1}{n}\sum^{n}_{i=1}(Y_i(1)-Y_i(0))
\end{align*}
which is the difference in the potential outcomes averaged across the $n$ units in the sample $S$. SATE could vary from sample to sample if the individual treatment effect $Y_i(1)-Y_i(0)$  is not constant. If $S$ is sampled randomly from the super-population, SATE is an unbiased estimate for PATE. 

Similarly, we assume the number of treated subjects in the population and sample are $N_1$ and $n_1$. We can define the population and sample treatment effect among the treated as
\begin{align*}
	\tau_{PATT}&=\mathbb{E}(Y(1)-Y(0)|A=1) \\
	&=\frac{1}{N_1}\sum^N_{i=1} A_i \big ( Y_i(1)-Y_i(0) \big)
\end{align*}
\begin{align*}
	\tau_{SATT}&=\mathbb{E}(Y(1)-Y(0)|A=1,S=1) \\
	&=\frac{1}{n_1}\sum^{n_1}_{i=1} A_i \big ( Y_i(1)-Y_i(0) \big)
\end{align*}

Under the assumption that individual treatment effects are constant, i.e., $Y_i(1)-Y_i(0) = \tau$,  $\tau_{PATE}=\tau_{SATE}=\tau_{PATT}=\tau_{SATT}$.  Under SUTVA and the conditional ignorable treatment assignment assumption, we can identify PATE with observed outcome. 
\begin{align*}
	\tau_{PATE}&=\mathbb{E}(Y_i(1)-Y_i(0)) \\
	&=\mathbb{E}(\mathbb{E}(Y_i(1)-Y_i(0)|\bm{X}_i)) \\
	&=\mathbb{E}(\mathbb{E}(Y_i|A_i=1,\bm{X}_i)-\mathbb{E}(Y_i|A_i=0,\bm{X}_i))
\end{align*}
The relationship between $Y$, $A$, and $\bm{X}$ was previously described using the following linear model \cite{king_why_2019}:
\begin{align}
	Y_i =\mathbb{E}(Y_i|A_i,\bm{X}_i)+\epsilon_i=\beta_0+\beta_1 A_i+g(\bm{X}_i)+\epsilon_i
	\label{eq_1}
\end{align}
where $\epsilon_i$ is a random error and $\mathbb{E}(\epsilon_i)=0$, $\beta_1$ represents the conditional exposure effect. $g(\cdot)$ is some arbitrary function. When the effects of $\bm{X}_i$ on $Y_i$ are linear additive, $g(\bm{X}_i)=\bm{\beta}_2 \bm{X}_i$, where $\bm{\beta}_2$ represents the $p-$dimensional vector of regression coefficients for $\bm{X}$. It follows that $\beta_1=Y_i(1)-Y_i(0)$ and $\tau_{PATT}=\tau_{SATE}=\tau_{PATT}=\tau_{SATT}=\beta_1$. King and Nielsen \cite{king_why_2019} stated that their causal interest lies in either SATE or SATT. However, model (\ref{eq_1}) implies a constant individual treatment effect. In this context, there is no distinction between population and sample causal estimands, nor between ATT and ATE. We adopted the same setting primarily to demonstrate that covariate imbalance in PSM occurs by chance and does not bias effect estimation, contrary to the findings of the prior study. We are not evaluating a novel estimation approach in more general settings. Finally, we focus on estimating the population effect, rather than the sample treatment effect, within our simulation design.

\subsection{Propensity Score}
\label{sec23}
The propensity score, denoted as $e(\bm{X})$, is formally defined as the conditional probability of receiving an active treatment given the baseline covariates $\bm{X}$. It serves as a summary score of $\bm{X}$. The propensity score 
$e(\bm{X})$ has two key properties:

\begin{Properties}
	\item (\textbf{Balancing score}) The propensity score $e(\bm{X})$ balances the distribution of $\bm{X}$ between the treatment groups:
	\begin{align*}
		A \independent \bm{X} | e(\bm{X})
	\end{align*}
	
	When pairing two subjects, denoted as $i$ and $j$, where one is treated and the other is untreated, such that $A_{ki} + A_{kj} = 1$, and they possess identical propensity scores in the $k$th matched pair, it's possible for these two subjects to exhibit different values for the observed covariates, $\bm{x}_{ki} \neq \bm{x}_{kj}$, despite having precisely the same propensity scores, $e(\bm{x}_{ki}) = e(\bm{x}_{kj})$. However, this discrepancy or mismatch in $\bm{X}$ within each matched pair can be attributed to chance and therefore cannot predict the treatment assignment \cite{rosenbaum_design_2020}. The crucial aspect of these within-pair mismatches is that the disparity in covariate values between treated and untreated subjects can fluctuate randomly, resulting in both positive and negative differences from one matched pair to another, occurring with equal frequency. In situations where the number of matched pairs is small, we can anticipate moderate imbalance between the treated ($A=1$) and untreated ($A=0$) groups. However, as the number of matched pairs increases, the between-group imbalance becomes negligible. This distinctive phenomenon of PS can also be elucidated within the conventional regression framework \cite{wan_interpretation_2021}. Confounding bias emerges when confounders, which are correlated with both the outcome and treatment variables, are excluded from the regression model of equation (\ref{eq_1}). By adjusting for PS rather than directly including confounders as covariates, confounders are effectively decomposed into two components: the PS itself and a residual term. Conditioning on the PS, this residual term becomes orthogonal to the treatment variable and can be considered as random noise. Omitting such random noise no longer biases the estimation of treatment effect in a regression model.
	
	It's important to note that successfully balancing the observed variables $\bm{X}$ by matching on $e(\bm{X})$ doesn't guarantee the balance of unmeasured variables. In practical applications, it is common to utilize the rule of thumb that considers a standardized mean difference (SMD), in absolute value, not exceeding 0.1 as a criterion to evaluate the balance of baseline variables individually between the treated and untreated groups. SMD of a confounder $X_j, j=1,2,\cdots, p$, is defined as:
	\begin{align*}
		d_j=\frac{\bar{X}_{1j}-\bar{X}_{0j}}{\sqrt{\frac{S^2_{1j}+S^2_{0j}}{2}}} 
	\end{align*}
	,where $\bar{X}_{1j}$, $\bar{X}_{0j}$, $S^2_{1j}$, and $S^2_{0j}$ are the sample means and sample variances of $X_j$ in treated and untreated groups.  SMD can capture both direction and size of the between-group imbalance in $X_j$.

	\item (\textbf{Strongly ignorable treatment assignment (SITA)}) If $A$ is unconfounded given $\bm{X}$, then $A$ is unconfounded given $e(\bm{X})$. Formally,
	\begin{align*}
		\{ Y(1),Y(0) \} \independent A | \bm{X} \rightarrow 	\{ Y(1),Y(0) \} \independent A | e(\bm{X})
	\end{align*}
	SITA requires that there are no unmeasured confounding variables and that there should be sufficient overlap in the propensity scores between the treated and untreated groups. If it suffices to match on $\bm{X}$ for a matching design, it also suffices to match on $e(\bm{X})$. Matching on a single variable $e(\bm{X})$ is more practical than matching on $\bm{X}$ when $\bm{X}$ is high-dimensional. 
\end{Properties}

In summary, two properties of PS ensure that any imbalance in baseline confounders between two comparison groups in an exactly matched PSM design occurs by random chance and cannot result in residual confounding that can lead to biased estimation of treatment effect. Conversely, if covariate imbalances are systematic rather than random, property 2 will not hold due to residual confounding. 

Finally, we assume $e(\bm{X})$ takes the following logit form:
\begin{align}
	e(\bm{X})=P(A=1|\bm{X})=\frac{e^{\alpha_0+\bm{\alpha_1} \bm{X}}}{1+e^{\alpha_0+\bm{\alpha_1} \bm{X}}}
	\label{eq_2}
\end{align}
where $\bm{\alpha_1}$ is the $p-$ dimensional coefficient vector. We will use formula (\ref{eq_2}) in simulation study.

\section{Results}

\subsection{The issues with the PSM paradox}
\label{sec3}

\subsubsection{Increase in imbalance}
Instead of confirming the balancing property of the PS in matching designs, King and Nielsen \cite{king_why_2019} demonstrated that as PSM approaches exact matching by eliminating the worst-matched pairs, the imbalance initially decreases. However, it subsequently increases beyond a certain threshold, deviating from continual improvement. Since this observation contradicts the balancing property of the PS, we need to carefully examine whether the metrics used in previous studies can adequately capture the chance imbalance. King and Nielsen \cite{king_why_2019} used the following imbalance metric:
\begin{align*}
	I(\bm{X})=\text{mean}_{i\in \{i\}}d(\bm{X}_i,\bm{X}_{j(i)})
\end{align*} 
Here $I(\bm{X})$ represents the average pairwise distance between treated subject $i$ with covariates $\bm{X_i}$, and the closest untreated subject $j$ with covariates $\bm{X_{j(i)}}$. $d(\cdot)$ is a distance function. For example, the Mahalanobis distance is a popular choice, which is defined as the following:
\begin{align*}
	d(\bm{X}_i,\bm{X}_{j(i)})=\sqrt{(\bm{X}_i-\bm{X}_{j(i)})^{'}\Sigma^{-1}(\bm{X}_i-\bm{X}_{j(i)})}
\end{align*}
where $\Sigma$  represents the sample
covariance matrix of the original data. Ripollone {\em et al}. \cite{ripollone_implications_2018} found a similar pattern of increasing imbalance after progressive pruning of worst matched pairs in real epidemiological data sets. Instead of using the average individual distance between matched subjects, Ripollone {\em et al} \cite{ripollone_implications_2018} measured imbalance using two different metrics: 1) the Mahalanobis distance between the covariate means in the treated and untreated groups, as follows:
\begin{align*}
	d(\bar{\bm{X}}_1,\bar{\bm{X}}_0)=\sqrt{(\bar{\bm{X}}_1-\bar{\bm{X}}_{0})^{'}\Sigma^{-1}(\bar{\bm{X}}_1-\bar{\bm{X}}_{0})}
\end{align*}
Here $\bar{\bm{X}}_1$ and $\bar{\bm{X}}_0$ are the vectors of covariate means in the treated and untreated groups. Larger Mahalanobis balance suggests worse covariate
balance. 2) $C$ statistic for the discriminatory
power of the logistic model predicting the treatment indicator in the matched data set. Higher $C$ statistic values suggest worse covariate balance.

The balancing property of the propensity score suggests that as the caliper size shrinks and PSM approaches exact matching, the distribution of $\bm{X}$ becomes the same between the treated and untreated groups. Any mismatches in $\bm{X}$ between treated and untreated subjects in matched pairs are random occurrences. However, we need to understand that the balancing of $\bm{X}$ between the two groups is a large sample property. To better understand this concept, consider a small randomized trial. In such cases, one might expect to observe significant mean differences in baseline covariates between treatment groups, even though treatments are randomly assigned. For instance, suppose a baseline variable $X$ in the treated and untreated groups in a randomized trial follow a normal distribution $\sim N(\mu, \sigma)$. In this scenario, the group means of $n$ subjects, denoted as $\bar{X}_1$ and $\bar{X}_0$, follow a normal distribution $\sim N(\mu, \frac{\sigma}{\sqrt{n}})$. SMD $d = \frac{\bar{X}_1 - \bar{X}_0}{\sigma} \sim N(0, \frac{2}{\sqrt{n}})$. When the sample size is small, the variance of $d$ becomes large, making it more likely to observe a significant imbalance between two groups even in a randomized trial. Thus, in a PSM design with a large number of matched pairs, we would expect the between-group imbalance to be negligible when averaging out these within-pair mismatches. In repeated samples with a finite sample size (e.g., simulation studies), chance imbalance implies that the between-group imbalance can be positive in one matched sample and negative in another. However, when averaged across all matched samples, it converges to zero. 

The issues with prior findings regarding the eventual increase in the between-group imbalance as PSM approaches exact matching are twofold: 
\begin{itemize}
	\item[(i)]The distance metrics used in previous studies \cite{king_why_2019,ripollone_implications_2018} fail to capture the inherent ``chance" aspect of observed imbalance in a PSM design. The Mahalanobis balance metric and $C$ statistics, employed in these studies, measure the absolute distance and consistently yield positive values without considering the directions of either within-pair or between group imbalances. Consequently, when the number of matched pairs increases, the within-pair Mahalanobis distances \cite{king_why_2019} cannot average towards zero, and when averaged over repeated samples, the between-group Mahalanobis distances and $C$ statistics \cite{ripollone_implications_2018} fail to converge towards zero as well. 
	\item[(ii)] Previous studies also demonstrated the PSM paradox by pruning the worst-matched pairs in a single dataset. However, balancing is a large sample property and the between-group imbalance in a PSM design tends to converge to zero with an increasing number of matched pairs, rather than a decreasing one from progressive pruning. When sample size is finite, this convergence towards zero occurs when averaged over repeated samples, not within a single sample. Instead, the noted rise in imbalance, as observed in prior studies \cite{king_why_2019,ripollone_implications_2018}, reflects the growing variability in chance imbalance as the sample size decreases through progressive pruning. Once PSM reaches the initial balance of $\bm{X}$ with an appropriate caliper (i.e., the point where PSM approximates a randomized design), further reduction in matched pairs results in a smaller sample size, thereby increasing the likelihood of large chance imbalances between two groups. Consequently, this leads to large Mahalanobis distances or $C$ statistics, as observed in previous studies \cite{king_why_2019,ripollone_implications_2018}.This trend is akin to small trials where the likelihood of observing significant baseline covariate imbalances is higher.
\end{itemize}

\subsubsection{Model dependence and bias}
\label{sec32}
As further revealed in the previous study \cite{king_why_2019}, increasing imbalance has consequences that include a rise in bias. However, chance imbalance does not predict treatment status and should not bias the estimation of PATE when PSM approaches exact matching, even when using the sample mean difference, one of the simplest estimators \cite{rosenbaum_central_1983}. So, where does this bias originate? It turns out that the bias observed in King and Nielsen's study \cite{king_why_2019} stems from an unconventional source - their choice of a generally biased estimator for the treatment effect. Consider a scenario in which an analyst has tried a set of different models $m_1, m_2, \cdots, m_J$ for estimating the treatment effect, resulting in corresponding estimates $\hat{\tau}_1,\hat{\tau}_2,\ldots, \hat{\tau}_J$ from each model. In such cases, researchers often opt for the maximum estimate among these, denoted as $\hat{\tau}_0 = \text{max}(\hat{\tau}_1,\hat{\tau}_2,\ldots, \hat{\tau}_J)$. As stated in the previous study\cite{king_why_2019}, this maximum coefficient $\hat{\tau}_0$ is typically biased, even when individual estimates are unbiased. 

King and Nielsen \cite{king_why_2019} provided some reasons behind the use of this biased estimator in assessing PSM: i) The data generation process and the true model are unknown, which may lead analysts to explore various models. In one of their examples, King and Nielsen fitted 512 different models for a simple PSM design with only two matching factors - an extreme case of post-matching analysis; ii) The diversity of estimates from all plausible models introduces model dependence, formally defined as the variance of estimates from all fitted models-$\hat{\sigma}^2 = \text{var}(\hat{\tau}_{1},\hat{\tau}_{2},\ldots, \hat{\tau}_{J})$\cite{king_why_2019}. Human choice can exacerbate model dependence, as researchers may select the maximum estimate, which is generally biased. However, it's important to note that this bias isn't due to confounding, which arises from systematic imbalances in confounders. 

There are two key issues with using cherry-picking to assess the model dependence of either a randomized design or a valid matching design that mimics randomization. First, it does not provide a valid assessment. For instance, consider two competing designs, $A$ and $B$, where all individual regression models produce unbiased estimates: $\{\hat{\tau}_{a1},\hat{\tau}_{a2},\ldots, \hat{\tau}_{aJ}\}$ and $\{\hat{\tau}_{b1},\hat{\tau}_{b2},\ldots, \hat{\tau}_{bJ}\}$, and $\mathbf{E}(\hat{\tau}_{aj})=\mathbf{E}(\hat{\tau}_{bj})=\tau,\forall j=1,2,\cdots, J$. Thus, unbiased estimation in both designs is independent of model specification, and any misspecified model can still yield unbiased estimates. Furthermore, we let $\hat{\tau}_{a0} = \text{max}(\hat{\tau}_{a1},\hat{\tau}_{a2},\ldots, \hat{\tau}_{aJ})$ and $\hat{\tau}_{b0} = \text{max}(\hat{\tau}_{b1},\hat{\tau}_{b2},\ldots, \hat{\tau}_{bJ})$, $\hat{\sigma}^2_{a} = \text{var}(\hat{\tau}_{a1},\hat{\tau}_{a2},\ldots, \hat{\tau}_{aJ})$ and $\hat{\sigma}^2_{b} = \text{var}(\hat{\tau}_{b1},\hat{\tau}_{b2},\ldots, \hat{\tau}_{bJ})$, if $\hat{\tau}_{a0} < \hat{\tau}_{b0}$ and $\hat{\sigma}^2_{a} < \hat{\sigma}^2_{b}$, Can we then reverse the previous conclusion and claim that design $A$ is less model-dependent than design $B$? Certainly not, as we know that model dependence is not an issue in either design. These differences primarily reflect variations in design efficiency, not model dependence. In a more efficient design, effect estimates are naturally less volatile. However, lower efficiency does not disqualify a design from being valid. For example, PSM is typically less efficient than other covariate matching methods, as it matches on a summary score rather than directly on covariates, using less information. However, it provides a practical solution to the curse of dimensionality, the key limitation of covariate-matching designs. Second, applying cherry-picking analysis to a matched design is not only biased but also unnecessary, as it overlooks one of the key benefits of matching designs - their resilience to model misspecification.

Regression analysis with confounders adjusted as covariates in the original unmatched data is commonly used by applied researchers to estimate the treatment effect due to its relative simplicity compared to a matching design. The primary issue associated with regression analysis is model dependence, where the outcome model must be fitted correctly (e.g., considering nonlinear forms of confounders and interaction terms). A misspecified model can lead to biased estimates. In contrast, when using matching design, one must navigate a complex matching algorithm and may have to discard valuable unmatched observations. If, after all these painstaking efforts, researchers continue to grapple with the challenge of selecting the correct outcome model amidst the exploration of numerous candidate models during post-matching analysis, they should question the benefit of using matching design over  regression adjustment in original data.

Researchers should be aware that matching serves as a nonparametric preprocessing tool. It has the ability to either eliminate, when exact matching is achieved, or reduce the reliance on correct model specifications in the post-matching analysis, as discussed by Ho et al.\cite{ho_matching_2007}. In cases of exact matching, a simple linear regression that includes only the treatment indicator, essentially representing the sample mean difference between the treatment and control groups, or a model adjusted for covariates, can both yield unbiased estimates. This is true even if these models happen to be misspecified \cite{guo_statistical_2023}. In situations where exact matching is not possible, the required adjustments are far less burdensome, less reliant on specific model assumptions than they would without matching. These informal claims find theoretical support in the work of Guo and Rothenh\"ausler \cite{guo_statistical_2023}. They also demonstrated that when exact matching is unattainable, additional linear regression adjustments become necessary.  Nonetheless, matching makes parametric analyses less sensitive to the correct model specification. These findings align with the earlier assertion that a well-designed matching process, coupled with subsequent regression adjustment, generally yields the least biased estimates \cite{rubin_use_1973}. Rather than experimenting with a multitude of different models for post-matching analysis, we can adopt a predefined approach by utilizing commonly employed misspecified models, such as the simple linear regression model or multiple linear regression, which includes linear terms for all matching factors. In the next section, we will assess the unbiasedness of these misspecified models using simulation.

\subsection{Simulation}
\label{sec5}

\subsubsection{Simulation design}
We conducted simulation studies to achieve the following objectives: (a) {\it{Assessing Systematic Imbalance}}: Our first inquiry aimed to determine whether there exists a systematic imbalance and a corresponding bias in effect estimation as we tightened the matching caliper size. (b) {\it{Evaluating Model Misspecification Sensitivity}}: Our second inquiry focused on assessing whether PSM reduces sensitivity to model misspecification. In order to generate the simulation data, we followed the methodology used in our previous studies \cite{wan_matched_2019, wan_evaluation_2018}, as detailed below: 
\begin{itemize}
	\item[(i)] We created two coefficient vectors, $\bm{\beta}_2$ and $\bm{\alpha}_1$, each containing five elements in the outcome and treatment models (equations (\ref{eq_1}) and (\ref{eq_2})). For $\bm{\beta}_2$, we initiated the elements of the coefficient vector by randomly sampling values from the range of 1 to 9. Subsequently, we normalized this coefficient vector to be a unit vector. The sign of each element was determined using a Bernoulli distribution with a probability of 0.5. Finally, we set $\bm{\beta}_2$ equal to $k$ multiplied by its normalized factor, with the value of $k$ being fixed at 1.2. The same procedure was then repeated to generate $\bm{\alpha}_1$, with $\bm{\alpha}_1$ set to 1 multiplied by the normalized vector. Among all generated pairs of coefficients, we specifically selected two pairs of coefficients with their the $sine$ distances falling within the intervals [0, 0.2] and (0.8, 1] respectively (Details in supplemental table 1 in Appendix 2).  The $sine$ distance measures the dissimilarity between two vectors, and it ranges from 0 to 1, with a larger value indicating a greater dissimilarity between the vectors \cite{wan_interpretation_2021}. When the $sine$ distance falls within the range of (0.8, 1], it signifies that there is a weak within-pair correlation among the matched subjects. In this context, the PSM design closely resembles a completely randomized design. Conversely, when the $sine$ distance is within the range of [0, 0.2], it suggests that the within-pair correlation among matched subjects is strong. In such instances, the PSM design approximates a blocked randomized design. Our approach intends to avoid the extreme results from using proportional coefficient sets of $\bm{\alpha}_1$ and $\bm{\beta}_2$ in prior studies \cite{wan_cautionary_2024}.
	
	\item[(ii)] For every pair of $\bm{\beta}_2$ and $\bm{\alpha}_1$, we generated five independent confounding variables, denoted as $X_1, X_2, \ldots, X_5$, from a normal distribution with mean 0 and standard deviation 1, each with a sample size of $n = 1500$. The treatment variable $A$ was created using the treatment model (\ref{eq_2}), with the intercept $\alpha_0$ set to -0.9. Thus, approximately 30\% of the simulated subjects received the treatment. The outcome variable $Y$ was generated using the linear model outlined in Equation (\ref{eq_1}), incorporating an error term $\sim N (0, 1)$. $\beta_1$ was set at 0.5 in section ``Imbalance and bias" and 1 in section ``Model dependence".
	
	\item[(iii)]  In each simulated data set, we computed the propensity score for every subject using a logistic regression model. We then applied a nearest-neighborhood matching algorithm to pair each treated subject with a control subject based on the logit of the propensity score. This matching was performed without replacement, utilizing a caliper width equal to $c$ times the standard deviation of the logit propensity score, where $c$ was selected from the set of values $\{20, 1, 0.2, 0.02, 0.002, 0.0002\}$.	Matching algorithm was implemented in $\texttt{R/MatchIt}$.
	
\end{itemize}

\subsubsection{Imbalance and bias}
\label{sec511}
At each matching caliper size, we calculated various metrics for assessing imbalance, including the number of the matched pairs, SMD of confounder $X_3$, and the multivariate Mahalanobis distance between group means of all confounders in the treated and untreated groups. This analysis was conducted on 5000 randomly generated samples, with the same calculations repeated in each sample. Subsequently, we averaged these measurements across the 5000 replicates. Additionally, we computed the proportion of SMDs of $X_3$ that exceeded 0.1 in the absolute values. We anticipated that the averaged SMD of $X_3$ would effectively capture any chance imbalance and thus converge toward 0 after the initial balancing was achieved. In contrast, we anticipated that the average of Mahalanobis metrics would initially decrease until a caliper size of 0.2 $\times$ the standard deviation of the logit propensity score was reached, after which they would increase.  The proportion of absolute SMDs of $X_3$ larger than 0.1 was also expected to follow this trend. These increasing trends reflect the increasing likelihood of observing substantial chance imbalances due to a continued decrease in sample size after the initial balancing was established.

In addition to calculating these imbalance metrics at each caliper size, we performed regression analyses. Three different models were fitted: an unadjusted model with the treatment indicator $A$ (referred to as ``$\mathcal{M}(A)$"), a multiple regression model including $A$, linear terms of $X_4$, and $X_5$ (referred to as the ``$\mathcal{M}(A,X_4,X_5)$"), a model including $A$, linear terms of $X_1$ through $X_5$ (referred to as the ``$\mathcal{M}(A,X_1,X_2,X_3,X_4,X_5)$"), and finally, we extracted estimates of the treatment effect from each model, along with model-based standard errors and robust sandwich standard errors. These effect estimates and variance estimates were then averaged across 5000 replicates. Additionally, we computed the empirical variance estimate for each effect estimator by calculating the variance of 5000 effect estimates. We anticipate that $\mathcal{M}(A,X_1,X_2,X_3,X_4,X_5)$ would outperform both $\mathcal{M}(A)$ and $\mathcal{M}(A,X_4,X_5)$, displaying the smallest bias and variance estimates. When considering the commonly recommended caliper size of 0.2 $\times$ the standard deviation of the logit propensity score, we expected that even though both $\mathcal{M}(A)$ and $\mathcal{M}(A,X_4,X_5)$ were misspecified, they will still produce nearly unbiased effect estimations.

\subsubsection{Model dependence}
\label{sec512}
We used a new set of coefficients and a more complex outcome model that incorporates quadratic and interaction terms to generate the outcome $Y$  (refer to the Appendix 1 for details). Within the framework of this complex outcome model, we evaluated the performance of misspecified models for effect estimation. Two different models were employed: $\mathcal{M}(A)$ and $\mathcal{M}(A,X_1,X_2,X_3,X_4,X_5)$. We then extracted estimates of the treatment effect from each model, along with model-based standard errors. These effect estimates and variance estimates were subsequently averaged across 5000 replicates. Additionally, we computed the empirical variance estimate for each effect estimator by calculating the variances of the 5000 effect estimates. We anticipate that with a caliper size of 0.2 times the standard deviation of the logit propensity score, both $\mathcal{M}(A)$ and $\mathcal{M}(A,X_1,X_2,X_3,X_4,X_5)$ would yield nearly unbiased effect estimates. To examine the claim that even inexact matching can still make parametric analyses less sensitive to model misspecification, we also fitted $\mathcal{M}(A,X_1,X_2,X_3,X_4,X_5)$ in unmatched data and computed the averaged estimate of $\beta_1$.

\subsubsection{Simulation results}

\begin{figure}[h!]
	\centering
	\includegraphics[width=\linewidth]{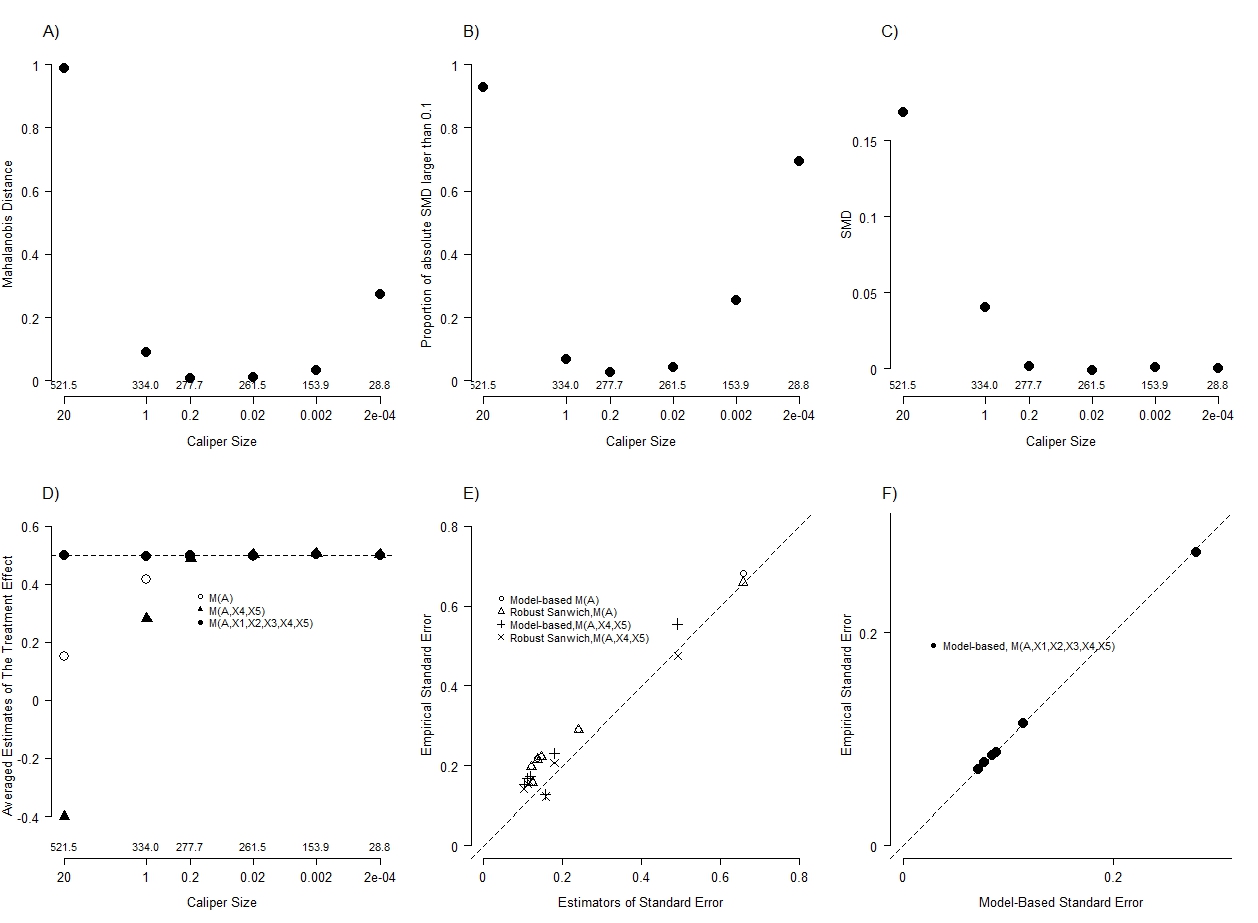}
	\caption{Imbalance and Bias for the $sine$ distance $>0.8$. A) The trend of the mahalanobis distance with shrinking caliper size; B) The trend of the proportion of absolute SMD of $X_3$ larger than 0.1; C) The trend of SMD of $X_3$; D) The trend of estimators of PATE; E) The concordance between empirical standard error and model-based/Robust Sandwich estimators for $\mathcal{M}(A)$ and $\mathcal{M}(A,X_4,X_5)$; F)The concordance between empirical standard error and model-based estimators for $\mathcal{M}(A,X_1,X_2,X_3,X_4,X_5)$ }
	\label{fig1}
\end{figure}

Figure \ref{fig1} illustrates the simulation results for imbalance and biases in the scenario involving low within-pair correlation, where the $sine$ distance exceeds 0.8 and PSM approximates a CRD. Firstly, as the caliper size diminishes, the Mahalanobis distance initially decreases until it reaches the optimal caliper size ($0.2 \times$ the standard deviation of logit PS), and then increases as the caliper size continues to shrink, and the sample size decreases (see Figure \ref{fig1}A). The proportion of absolute SMD greater than 0.1 for $X_3$ follows a similar pattern (Figure \ref{fig1}B). SMD of $X_3$ decreases until reaching the optimal caliper size and then stabilizes near zero thereafter (Figure \ref{fig1}C). This confirms that metrics like the Mahalanobis distance only reflects the increasing variability in chance imbalance and a higher likelihood of observing larger chance imbalance as the sample size decreases. Only metrics that retain the direction of differences, such as SMD, can accurately measure this chance imbalance. Secondly, we can also observe that $\mathcal{M}(A)$ and $\mathcal{M}(A,X_4,X_5)$ produce biased estimates until confounders are balanced at the optimal caliper size, and PSM approximates a randomized design (Figure \ref{fig1}D). Given that the true outcome model includes linear terms for five confounders, $\mathcal{M}(A,X_1,X_2,X_3,X_4,X_5)$ consistently produces unbiased results at all caliper sizes. When either the regression model is correctly specified or matching is performed correctly (even with incorrect models), we can consistently obtain unbiased results (Figure \ref{fig1}D). Lastly, it's worth noting that the model-based standard errors for both $\mathcal{M}(A)$ and $\mathcal{M}(A,X_4,X_5)$ do not align precisely with their empirical standard errors, and robust sandwich estimators do not show significant improvement (Figure \ref{fig1}E). As expected, the model-based standard errors from $\mathcal{M}(A,X_1,X_2,X_3,X_4,X_5)$ align well with its empirical standard errors (Figure \ref{fig1}F).

\begin{figure}[h!]
	\centering
	\includegraphics[width=\linewidth]{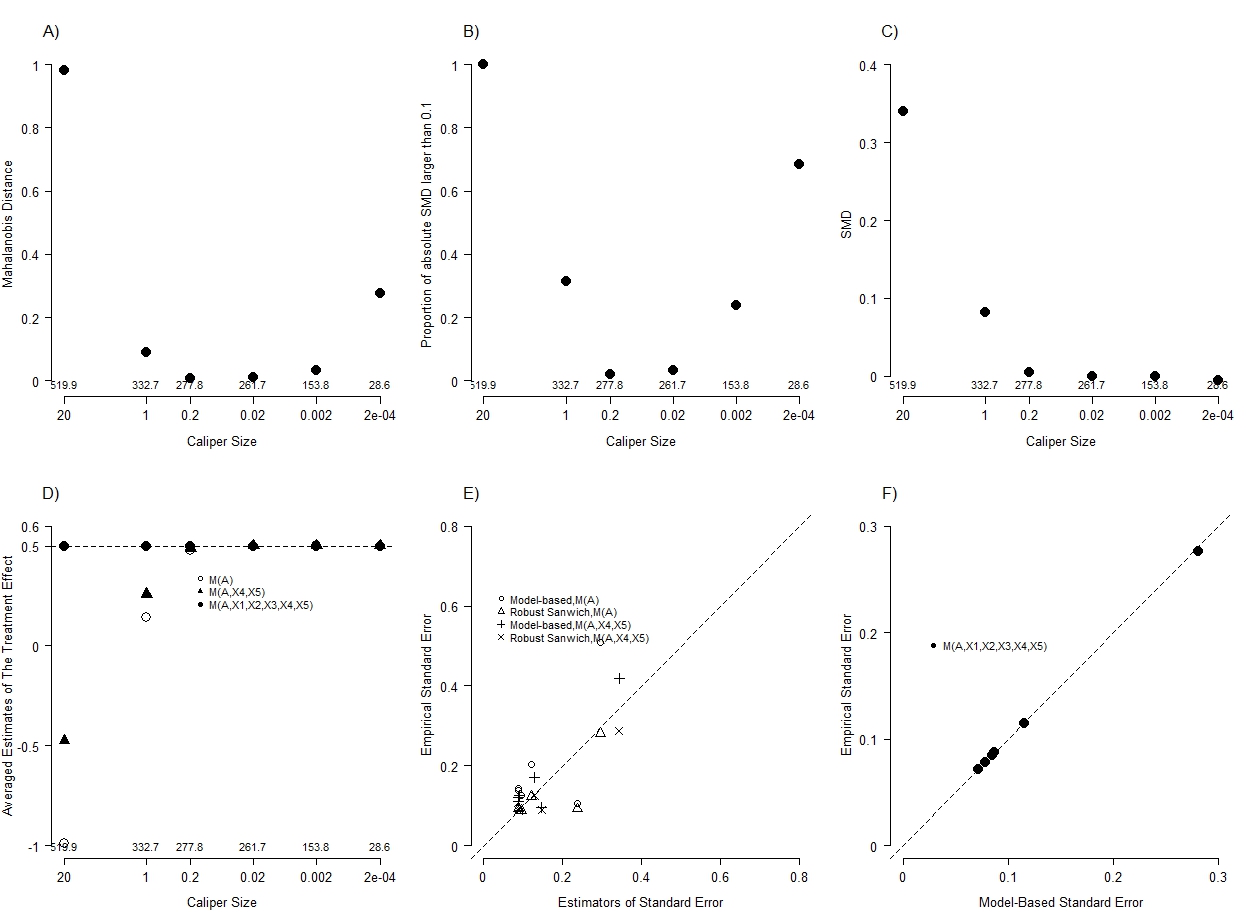}
	\caption{Imbalance and Bias for the $sine$ distance $<0.2$. A) The trend of the mahalanobis distance with shrinking caliper size; B) The trend of the proportion of absolute SMD of $X_3$ larger than 0.1; C) The trend of SMD of $X_3$; D) The trend of estimators of PATE; E) The concordance between empirical standard error and model-based/Robust Sandwich estimators for $\mathcal{M}(A)$ and $\mathcal{M}(A,X_4,X_5)$; F)The concordance between empirical standard error and model-based estimators for $\mathcal{M}(A,X_1,X_2,X_3,X_4,X_5)$ }
	\label{fig2}
	
\end{figure}

Figure \ref{fig2} illustrates the simulation results for imbalances and biases in a scenario where the within-pair correlation is high (i.e., the $sine$ distance is less than 0.2 and PSM approximates a Blocked Randomized Design). We observe a consistent pattern of imbalance and bias across different aspects (Figure \ref{fig2} A-D). Both $\mathcal{M}(A)$ and $\mathcal{M}(A,X_4,X_5)$ exhibit bias until the caliper size reaches the optimal level. In contrast, the correctly specified $\mathcal{M}(A,X_1,X_2,X_3,X_4,X_5)$ remains unbiased throughout the process. Notably, robust sandwich estimates for $\mathcal{M}(A)$ and $\mathcal{M}(A,X_4,X_5)$ perform better than model-based estimates after the caliper size reaches the optimal level (Figure \ref{fig2} E and Table 2 in Appendix 2). As expected, the model-based standard errors of $\mathcal{M}(A,X_1,X_2,X_3,X_4,X_5)$ closely align with their empirical standard errors (Figure \ref{fig2}F).

\begin{figure}[h!]
	\centering
	\includegraphics[width=\linewidth]{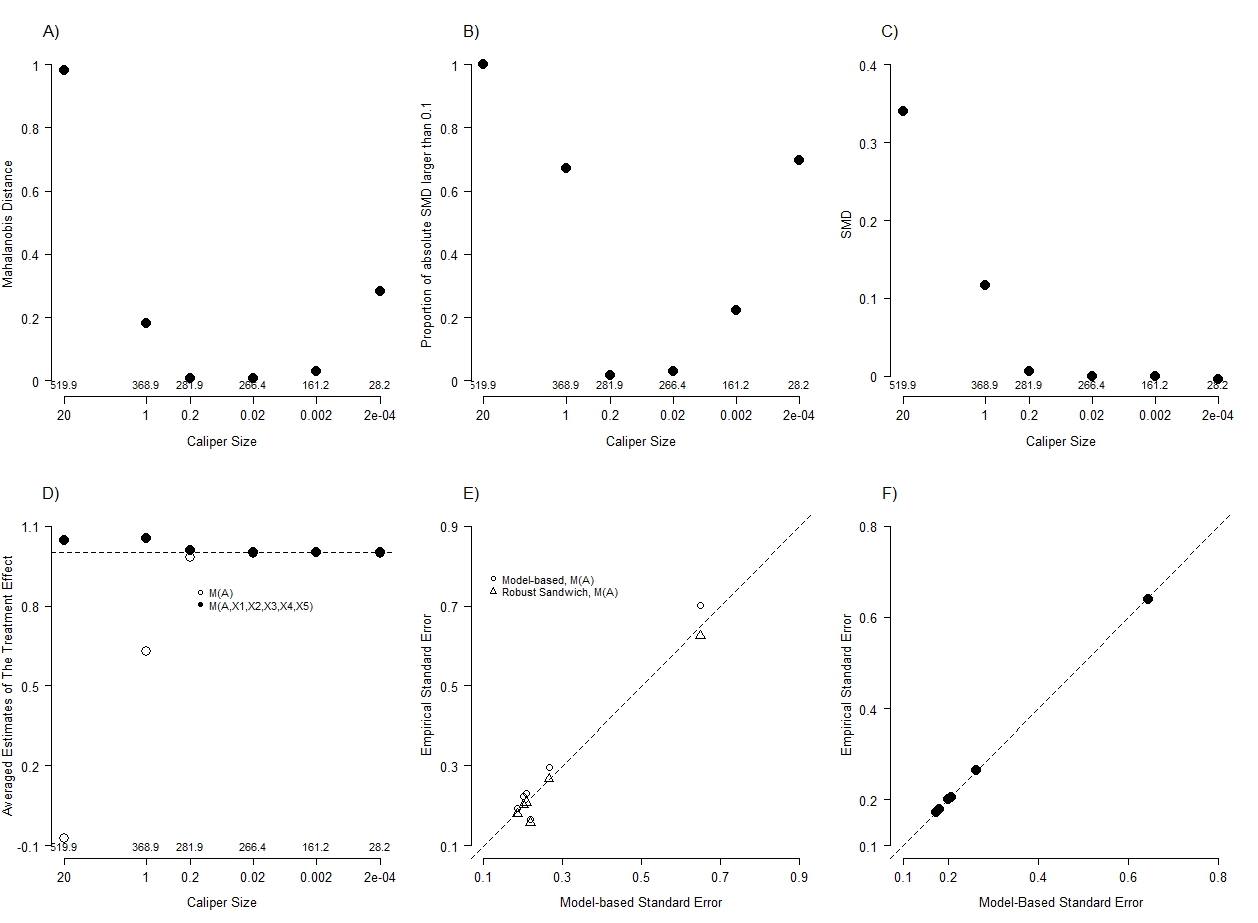}
	\caption{Sensitivity to Model Dependence. A) The trend of the mahalanobis distance with shrinking caliper size; B) The trend of the proportion of absolute SMD of $X_3$ larger than 0.1; C) The trend of SMD of $X_3$; D) The trend of estimators of PATE; E) The concordance between empirical standard error and model-based estimators for $\mathcal{M}(A)$; F)The concordance between empirical standard error and model-based estimators for $\mathcal{M}(A,X_1,X_2,X_3,X_4,X_5)$ }
	\label{fig3}
	
\end{figure}

Figure \ref{fig3} illustrates the sensitivity of PSM to model misspecification in our simulation results. We observe a consistent pattern in imbalance metrics (Figure \ref{fig3}A-C). Both misspecified models show bias until the matching caliper size reaches its optimal level (Figure \ref{fig3}D). Notably, model-based standard errors for $\mathcal{M}(A,X_1,X_2,X_3,X_4,X_5)$ closely match empirical standard errors across all caliper sizes. However, this alignment is poor for $\mathcal{M}(A)$, although the robust sandwich variance estimator improves the alignment at and after the optimal caliper size (Figure \ref{fig3}F and Table 3 in Appendix 2). Moreover, the averaged estimate of $\beta_1$ obtained using $\mathcal{M}(A,X_1,X_2,X_3,X_4,X_5)$ in the unmatched dataset is 1.37, indicating a considerably higher bias compared to when the same model is applied in a poorly matched PSM design, employing a large caliper size of 20 $\times$ the standard deviation of the logit PS (Figure \ref{fig3}F). This reaffirms the previous conclusion \cite{guo_statistical_2023} that matching, even an imperfect one, can effectively reduce the sensitivity of parametric modeling to model misspecification.

In summary,  in line with the balancing score property of PS, the observed imbalance in PSM primarily arises due to chance, which cannot be adequately captured by absolute distance metrics like the Mahalanobis distance. This chance-driven imbalance eventually averages to zero when the matching caliper size is optimal, thereby not biasing effect estimation when using misspecified models. The combination of matching design and post-matching regression analysis can be seen as a double-robust estimation procedure \cite{ho_matching_2007}, and PSM reduces sensitivity to model misspecification in post-matching analysis. We find that model-based inference using $\mathcal{M}(A,X_1,X_2,X_3,X_4,X_5)$ appears valid in PSM, regardless of whether it is misspecified or not. It's essential to note that selectively choosing the best result from a multitude of regression models is not only biased but also unnecessary.

\section{Discussion}
\label{sec6}
In this study, we have successfully validated the theoretical properties of PS in the context of matching designs. Specifically, when PSM approaches exact matching, it effectively balances confounding variables between comparison groups, and any observed imbalances are merely due to chance. Additionally, our findings confirm that PSM design mitigates the sensitivity to model misspecification in post-matching analysis, a characteristic described by Guo and Rothenh\"ausler for a matching design \cite{guo_statistical_2023}. Both a simple group mean difference and regression adjustment using linear terms of matching factors can accurately estimate PATE. Our findings stand in contrast to the previously identified PSM paradox \cite{king_why_2019}. This paradox suggests that model dependence and statistical bias should increase as units are pruned. We conclude that this discrepancy primarily arises from the use of inappropriate metrics for assessing imbalance and bias in the previous study \cite{king_why_2019}. Consequently, there is no valid concern that should deter us from employing PSM in comparative effectiveness research. 

Differing from other highly cited studies that emphasize PATE \cite{austin_comparison_2014}, King and Nielsen \cite{king_why_2019} have stated that SATE is the causal interest. SATE represents the treatment effect exclusively within the context of the available sample data, rather than being applicable to the broader target population. However, the inclusion of SATE in their work appears primarily intended to facilitate a conceptual understanding of the PSM paradox. In other words, if there is an imbalance within a sample, and the causal effect is specific to that particular sample, it may lead to a confounding bias when estimating this sample-specific quantity. Upon closer examination of their simulation design, it becomes evident that King and Nielsen are, in fact, assessing PATE. In their simulation studies, the causal parameter is represented by the coefficient of the treatment indicator in an additive model that does not include treatment-by-confounder interactions. This is a conventional method for representing a homogeneous population treatment effect, and they draw different random samples to calculate the average effect estimate. This approach aligns with the typical methodology for computing the expected value of an effect estimator for PATE, similar to what we have done in this study and other studies \cite{austin_comparison_2014}. Therefore, it is important to note that King and Nielsen have not formally demonstrated the bias in estimating SATE in PSM design. Even if such a demonstration was made, it would not be sufficient grounds to dismiss the utility of PSM because PSM was initially proposed to target PATE and has been substantiated through simulation studies in this context \cite{austin_methods_2009, austin_comparison_2014}.

We have also demonstrated that the prior findings concerning the eventual increase in imbalance as units are progressively pruned \cite{king_why_2019,ripollone_implications_2018}, a significant consequence of the PSM paradox, can be attributed to the inappropriate imbalance metrics that were employed. As extensively discussed by Rosenbaum \cite{rosenbaum_design_2020}, the covariate values of treated and untreated subjects matched on the same propensity score often exhibit different covariate values, which occur by chance and can swing in either direction. Consequently, when the number of matched pairs is substantial, these differences tend to average towards zero in PSM design. When matched samples are finite, the imbalances, when averaged across all matched samples, converge toward zero, or in other words, the expected value of imbalance is zero. Therefore, attempting to utilize the averaged pairwise Mahalanobis absolute distance in covariate space between comparison groups as a measure of imbalance represents a misinterpretation of the balancing property inherent in PS.

Chance imbalances doesn't predict the treatment status and should not introduce bias when estimating PATE \cite{rosenbaum_design_2020}. This issue has relevant discussion in randomized design setting, where unadjusted statistical tests for comparing treatment arms remain valid even in the presence of chance imbalances \cite{senn_seven_2013}. King and Nielsen \cite{king_why_2019} showed increased model-dependence and statistical bias connected with the PSM paradox using a cherry-picking estimation procedure. However, we argue that evaluating model dependence of PSM based on biased cherry-picking procedure is not appropriate.  As shown in our simulation and other studies \cite{ho_matching_2007,guo_statistical_2023}, PSM reduces the sensitivity of model misspecification because we can fit a well-considered model for estimating the treatment effect with satisfactory precision and efficiency, even if the chosen model could be misspecified. The key advantage of PSM, as a matching design, is that it frees us from the need to explore all possible models and search for the best results in post-matching analysis, which is a biased practice that should be avoided. This practice can lead to the most dramatic effect estimate even in randomized studies \cite{tsiatis_covariate_2008}. The reduction on model dependence by PSM can be explained by the fact that matching can balance any function of $\bm{X}$, making $A$ and any function of $\bm{X}$  approximately orthogonal in the matched sample. Thus, the inclusion or exclusion of a nearly orthogonal predictor has negligible effects on the other regression coefficients based on least-square theory \cite{guo_statistical_2023}. This resembles similar practice in analyzing a RCD. We can perform unadjusted analysis and also adjusted regression with gains in efficiency. However, what variables should enter into model and what forms they should take must be determined at the design stage. 

Moreover, valid inference by a misspecified regression model not only relies on the unbiasedness of its effect estimator but also on validity of its variance estimator. We found that the model including the treatment indicator and linear terms of matching factors provides valid model-based inference, regardless of whether it is misspecified or not, which provides some empirical support to the conclusion from \cite{guo_statistical_2023}. Our simulation results also support the previous recommendation on combining matching with regression in post-matching analysis \cite{guo_statistical_2023,rubin_use_1973}. This hybrid approach can be considered as a double-robust approach \cite{ho_matching_2007}. When matching is not exact, correct outcome modeling in post-matching analysis can lead to unbiased estimate. When matching is exact, linearly adjusted regression model, even potentially misspecified, can still lead to unbiased estimate. Future research on this topic is necessary.

\section{Conclusions}

Like any other statistical methods, PSM is not a universal solution. PSM does have its advantages and limitations. For instance, PSM offers a convenient way to address the curse of dimensionality problem in matching design. However, it may falter when confronted with significant unmeasured confounding. It's also crucial to note that achieving unbiased estimation of the treatment effect in PSM requires the correct specification of the propensity score model \cite{lenis_measuring_2018}, even though matching design is resilient against misspecification in outcome modeling. Nevertheless, it's important to recognize that the PSM paradox should not overly concern researchers.

\bmhead{Supplementary information}

Supplementary Material 1.

\section*{Data availability }
No datasets are used in this study

\section*{Ethics approval and consent to participate}
Not applicable

\section*{Consent for publication}
Not applicable

\section*{Competing interests}
The author declares no competing interests.

\section*{Funding}
None

\section*{Acknowledgements}
None

\section*{Authors' contributions}
FW was responsible for the conception of the paper's ideas, the design and execution of the simulations, and the drafting of the manuscript.




%\end{appendices}


\bibliography{sn-bibliography}% common bib file

\clearpage

\newpage

\end{document}